\documentclass[prx,reprint,nofootinbib,amssymb,superscriptaddress]{revtex4-2}

\usepackage{notes2bib}
\usepackage{natbib}
\usepackage{gensymb}
\usepackage{amsthm}
\usepackage{amsmath}
\usepackage{color}
\usepackage{bm}
\usepackage{amsmath,amssymb}
\usepackage{amsfonts}
\usepackage{dsfont}
\usepackage{graphicx}
\usepackage{float}
\usepackage{subfigure}
\usepackage{verbatim}
\usepackage{amsfonts}
\usepackage{bbold}
\usepackage{dcolumn}
\usepackage{bm}
\usepackage[dvipsnames]{xcolor}
\usepackage{listings}
\definecolor{blueprl}{RGB}{46,48,146}
\usepackage[pdftex,colorlinks=true,urlcolor=blueprl,citecolor=blueprl,linkcolor=blueprl]{hyperref}
\usepackage{times}
\usepackage{color}
\usepackage[dvipsnames]{xcolor}

\usepackage{graphicx}
\usepackage{amsmath}
\usepackage{enumitem}
\usepackage{amssymb}
\usepackage{array}
\newcolumntype{L}[1]{>{\raggedright\let\newline\\\arraybackslash\hspace{0pt}}m{#1}}
\newcolumntype{C}[1]{>{\centering\let\newline\\\arraybackslash\hspace{0pt}}m{#1}}
\newcolumntype{R}[1]{>{\raggedleft\let\newline\\\arraybackslash\hspace{0pt}}m{#1}}
\usepackage{mathrsfs}
\usepackage{appendix}
\usepackage{multirow}
\usepackage{dsfont}
\usepackage{tikz}
\usepackage{amsfonts}
\usepackage{amssymb}
\usepackage{pgflibraryarrows}
\usepackage{pgflibrarysnakes}
\usetikzlibrary{decorations.pathmorphing}
\usepgflibrary{arrows.meta}

\usepackage{subfigure}
\usepackage{bbold}
\newcommand{\D}{\mathrm{d}}

\newcommand{\dd}{\dagger}

\newcommand{\langlem}{\langle 0_M|} 
\newcommand{\ranglem}{|0_M\rangle}

\newcommand{\infint}{\int_{-\infty}^{+\infty}}
\usepackage{accents}
\usepackage{braket}

\newcommand{\vt}{\vphantom{\frac{1}{\sqrt{2}}}}
\newcommand{\non}{\nonumber}
\newcommand{\Mtx}{\mathbf{X}}

\graphicspath{{Figures/}}
\allowdisplaybreaks

\usepackage{xtab,afterpage}
\usepackage{hyperref}

\definecolor{blueblue}{RGB}{21,47,181}

\hypersetup{ linktoc=all,
    colorlinks, linkcolor={blueblue},
    citecolor={blueblue}, urlcolor={blueblue}
}

\begin{document}

\title{Unruh-deWitt detectors in quantum superpositions of trajectories}
%\title{Accelerated time-delay hides information through correlations in vacuum noise}

%\author{Daiqin Su and T.C.Ralph}
\author{Joshua Foo}
\email{joshua.foo@uqconnect.edu.au}
\author{Sho Onoe}
\affiliation{Centre for Quantum Computation and Communication Technology, School of Mathematics and Physics, The University of Queensland, St. Lucia, Queensland, 4072, Australia}
\author{Magdalena Zych}
\email{m.zych@uq.edu.au}
\affiliation{Centre for Engineered Quantum Systems, School of Mathematics and Physics, The University of Queensland, St. Lucia, Queensland, 4072, Australia}

\date{\today}

\begin{abstract}
{Unruh-deWitt detectors have been utilised widely as probes for quantum particles, entanglement and spacetime curvature. Here, we extend the standard treatment of an Unruh-deWitt detector interacting with a massless, scalar field to include the detector travelling in a quantum superposition of classical trajectories. We derive perturbative expressions for the final state of the detector, and show that it depends on field correlation functions evaluated locally along the individual trajectories, as well as non-locally between the superposed trajectories. By applying our general approach to a detector travelling in a superposition of two uniformly accelerated trajectories, including those with equal and differing proper accelerations, we discover novel interference effects in the emission and absorption spectra. These effects can be traced to causal relations between the superposed trajectories. Finally, we show that in general, such a detector does not thermalise even if the superposed paths would individually yield the same thermal state.}
\end{abstract}
\maketitle
%\vspace{10 mm}

\section{Introduction}
The Unruh-deWitt (UdW) detector is widely used as a probe of the foundational aspects of relativistic quantum fields and the structure of spacetime. The standard formulation of the model describes an idealised particle detector -- typically, a point-like two-level system -- that follows a classical worldline and whose internal states couple to the field \cite{unruh1984happens}. For example, consider the detector interacting with the massless scalar field in the Minkowski vacuum, and traversing a uniformly accelerated trajectory in spacetime with proper acceleration $a$. Unlike an inertial detector, which register no particles, the accelerated detector perceives a thermalised quantum state, radiating particles at the Unruh temperature, 
\begin{align}
    T_U &= \frac{a}{2\pi }.
\end{align}
This phenomenon is a manifestation of the Unruh effect, a prediction of relativistic quantum field theory that asserts that the experience of observers -- i.e. detectors -- interacting with quantum fields is frame-dependent \cite{unruh1976notes}. The utility and simplicity of the UdW detector model has facilitated its application to numerous related problems. Perturbative \cite{davies2002detection,schlicht2004considerations,louko2006often} and non-perturbative \cite{lin2007backreaction,brown2013detectors,bruschi2013time} approaches have been used to study entanglement dynamics and detection in settings such as non-inertial reference frames and expanding universes \cite{ ostapchuk2012entanglement,salton2015acceleration,kukita2017entanglement,martin2014entanglement,nambu2013entanglement,menicucciPhysRevD.79.044027}, and detector responses in curved spacetimes \cite{louko2008transition,hodgkinson2012static,hodgkinson2014static,ng2014unruh,birrell1984quantum} and higher-dimensional topologies \cite{langlois2006causal,hodgkinson2012often}. A list of further results can be found in \cite{hu2012relativistic}. 

While the UdW detector model has been devised as a probe for quantum particles, the quantum effects associated with its motion are just beginning to be explored. In particular, \cite{stritzelberger2019coherent} studies the absorption and emission of a UdW detector with a position degree of freedom described by a freely expanding wavefunction. In this work, we develop a general description of a single UdW detector coupling to a massless, scalar field and travelling in a \textit{quantum superposition of classical trajectories}. More specifically, we are interested in the response of such a detector when subjected to a combination of relativistic and quantum-mechanical effects. By initialising the detector in a quantum-controlled superposition of uniformly accelerated trajectories (parallel and anti-parallel accelerations, and trajectories sharing a Rindler horizon with differing proper accelerations), we discover the presence of novel interference dynamics in the emission and absorption spectra. These effects depend on the causal relations between the trajectories of the superposition, mediated by the non-local correlation functions evaluated along different trajectories.  

Such a model may engender new approaches for studying fundamental aspects of relativistic quantum field theory. For example, it provides an operational approach for studying the behaviour of quantum fields in \textit{quantum reference frames} (reference frames defined with respect to systems possessing quantum indeterminacy) \cite{giacomini2019quantum}, since the detector either does not have a well-defined spacetime location (parallel, anti-parallel accelerations) or proper acceleration (differing accelerations). In the latter case, one may ask whether the notion of a coherent superposition of temperatures can be meaningfully defined, given that the individual accelerations are associated with a unique Unruh temperature. Furthermore, the possibility of superposing the temporal order of the detector's interaction with the field may enable it to probe the causal structure of spacetime \cite{oreshkov2012quantum,zych2019bell,dimic2017simulating}. We comment on the former question in Sec.\ \ref{Superpose} while we propose the latter as a future direction in Sec.\ \ref{conclusion}. As we elaborate upon in Sec.\ \ref{conclusion}, quantum-controlled UdW detectors also unveil a deeper connection between coherently controlled quantum channels \cite{oi2003interference, ebler2018enhanced, chiribella2019quantum,abbott2018communication}, relativistic quantum information \cite{mann2012relativistic}, and quantum thermodynamics \cite{vinjanampathy2016quantum}. 

This paper is organised as follows: in Sec.\ \ref{density}, we review the UdW detector model coupling to a massless, scalar field, and apply it to a detector in an arbitrary superposition of relativistic -- i.e.\ classical -- trajectories. We then derive expressions for the conditional transition probability and instantaneous transition rate of the detector to second-order in perturbation theory. In Sec.\ \ref{Superpose}, we apply our formalism to a two-trajectory superposition of uniformly accelerated paths in parallel and anti-parallel motion, and those sharing a Rindler horizon but with differing proper accelerations. We conclude with some final remarks and directions for future research. Throughout, we use natural units $c = \hslash = k_B = 1$ and the metric signature $(-,+,+,+)$.

\section{Final Detector State, Transition Probabilities and Rates}\label{density}
\subsection{Unruh-deWitt Model}\label{udwmodel}
We begin by considering a two-level Unruh-deWitt detector initially in its ground state $|g\rangle$ and coupled to the real, massless scalar field, $\hat{\Phi}(\mathbf {x}(\tau))$, in (1+3)-dimensional Minkowski spacetime. Suppose that the field is also in its ground state, the Minkowski vacuum $\ranglem$. To initialise the detector in a trajectory superposition, we introduce a control degree of freedom, $i$, whose states $|i\rangle$ designate the individual paths which the detector takes. The state of the system can be expressed as
\begin{align}
    | \psi \rangle = | 0_M \rangle \otimes | g \rangle \underbrace{ \otimes \frac{1}{\sqrt{N}} \sum_{i=1}^N | i \rangle }_{| \mu \rangle } ,
\end{align}
where $\{ | i \rangle \}$ are orthogonal states of the control. Furthermore, we neglect any free dynamics of the control and assume that it is unaffected by measurements of the internal states of the detector such as its energy levels.

Now, the coupling of the detector to the field is described by the interaction Hamiltonian,
\begin{align}\label{ref:2}
\hat{H}_\mathrm{int.}(\tau)  &= \sum_{i=1}^N \hat{\mathcal{H}}_i \otimes |i\rangle\langle i |
\end{align}
where the individual terms
\begin{align}
    \hat{\mathcal{H}}_i (\tau) &= \lambda \eta(\tau) \sigma(\tau) \hat{\Phi}(\Mtx_i(\tau) ),
\end{align}
and $\lambda\ll 1$ is a weak coupling constant, $\eta(\tau)$ is a time-dependent switching function that governs the interaction, $\sigma(\tau) = \sigma^+ e^{i\Omega\tau} + \mathrm{h.c}$ is the interaction picture Pauli operator (with $\sigma^+ = |e\rangle\langle g|$) for the detector with energy gap $\Omega$ between the energy eigenstates $|g\rangle,|e\rangle$, and $\Mtx_i(\tau)$ is the worldline of the $i$th path of the superposition. The evolution of the initial system state from an initial time $\tau_i$ to final time $\tau$ can be obtained by perturbatively expanding the time evolution operator using the Dyson series,  
\begin{align}\label{eqn:dyson}
    \hat{U} &= \mathds{I} -  i\lambda  \int_{\tau_i}^{\tau_f} \D \tau \: \hat{H}_\mathrm{int.}(\tau) \nonumber \\
    & -\lambda^2 \int_{\tau_i}^{\tau_f} \D \tau \int_{\tau_i}^{\tau} \D \tau' \: \hat{H}_\mathrm{int.}(\tau) \hat{H}_\mathrm{int.}(\tau') + \: \mathcal{O}(\lambda^3)
\end{align}
where we have truncated the series beyond $\mathcal{O}(\lambda^2)$. The upper integration limit the second-order term enforces time-ordering of the Hamiltonians. $\hat{U}$ can be expressed as
\begin{align}
    \hat{U} &= \sum_{i=1}^N \hat{U}_i \otimes |i\rangle\langle i| ,
\end{align}
where
\begin{align}
    \hat{U}_i &= 1 - i \lambda \int_{\tau_i}^{\tau_f} \D \tau \: \hat{\mathcal{H}}_i(\tau) \nonumber \\
    & - \lambda^2 \int_{\tau_i}^{\tau_f} \D \tau \int_{\tau_i}^{\tau} \D \tau' \: \hat{\mathcal{H}}_i(\tau) \hat{\mathcal{H}}_i(\tau') + \mathcal{O}(\lambda^3)
\end{align}
are the contributions to $\hat{U}$ along the $i$th trajectory of the superposition. The time evolution of $|\psi\rangle$ is thus given by
\begin{align}
    \hat{U} |\psi\rangle  &= \frac{1}{\sqrt{N}} \sum_{i=1}^N \hat{U}_i \left( | 0_M \rangle \otimes | g \rangle \otimes | i \rangle \right)  ,
\end{align}
where the detector traversing the $i$th trajectory interacts locally with the field $\hat{\Phi}(\Mtx_i(\tau))$ along the worldline $\Mtx_i(\tau)$.

\subsection{Conditional Transition Probability}
We consider the conditional transition probability and instantaneous transition rate of the detector given that the control is measured in a superposition state, which for simplicity we take to be $|\mu \rangle = (1/\sqrt{N})\sum_{i=1}^N | i \rangle$. This measurement is assumed to take place in the asymptotic future, and describes a Mach-Zehnder type interference where the interferometric paths are associated with the trajectories of the detector. A generalised case, where the final state of the control is an arbitrary superposition state, is discussed in Sec.\ \ref{sec:general}. The final state of the detector-field system is given by $\langle \mu | \hat{U}| \psi\rangle = | \psi \rangle_\mathrm{FD}$,
\begin{align}
    |\psi\rangle_\mathrm{FD} &= \frac{1}{N} \sum_{i=1}^N \hat{U}_i \big( \ranglem \otimes |g\rangle \big) ,
\end{align}
with density matrix 
\begin{align}\label{10}
\hat{\rho}_\mathrm{FD} &= \frac{1}{N^2} \sum_{i,j=1}^N \underbrace{  \hat{U}_i \ranglem |g\rangle\langle g | \langlem \hat{U}_j^\dd }_{\hat{\rho}_{ij,\mathrm{FD}}} .
\end{align}
Using the series expansion Eq.\ (\ref{eqn:dyson}) and tracing over the final states of the field, the terms in the density matrix can be written as
\begin{align}\label{rhod}
    \mathrm{Tr}_\Phi \left[ \hat{\rho}_{ij,\mathrm{FD}} \right] &= \left( 1 - \frac{1}{2} ( P_{ii,E} - P_{jj,E}) \right) | g \rangle \langle g | + P_{ij,E} | e \rangle \langle e | ,
\end{align}
and the ground and excited state contributions are given by 
\begin{align}
    P_{ij,E} &= \lambda^2 \infint\D \tau_i \D \tau_i' \: \eta(\tau_i) \eta(\tau_i')  e^{-i\Omega (\tau_i - \tau_i')} W_{ji}(\Mtx_i(\tau_i), \Mtx_j'(\tau_j')) ,
\end{align}
where the Wightman functions 
\begin{align}
    W_{ji}(\Mtx_i(\tau_i), \Mtx_j'(\tau_j')) &= \langlem \hat{\Phi}(\Mtx_i(\tau_i) ) , \hat{\Phi}(\Mtx_j'(\tau_j' ))  \ranglem 
\end{align}
are evaluated along the worldlines $\Mtx_i(\tau), \Mtx_j'(\tau')$, the coordinates of which parametrise the field operators. We have also taken $\tau_0 \to -\infty$ and $\tau_f \to \infty$ for illustration. Tracing out the states of the field as usual, one obtains the reduced density matrix,
\begin{align}\label{eqn:density}
    \hat{\rho}_D &= \frac{1}{\mathcal{N}}\begin{pmatrix} 1- P_G & 0 \\ 0 & P_E \end{pmatrix} \simeq \begin{pmatrix} 1 - P_E & 0 \\ 0 & P_E \end{pmatrix} + \mathcal{O}(\lambda^4),
\end{align}
where $\mathcal{N}= 1 - P_G + P_E$ normalises the final state, conditioned upon measuring the control in $| \mu \rangle$, and 
\begin{widetext}
\begin{align}
    1 - P_G &= 1 - \frac{2\lambda^2}{N^2}\sum_{i=1}^N \infint\D \tau_i \infint \D \tau_i' \: \eta(\tau_i) \eta(\tau_i') e^{-i\Omega(\tau_i' - \tau_i')} W_{ii}(\Mtx_i(\tau_i), \Mtx_i'(\tau_i')), \vphantom{\infint} \\ \label{excitation}
    P_E &= \frac{\lambda^2}{N^2} \sum_{i=1}^N \sum_{j=1}^N \infint\D \tau_i \infint\D \tau_j' \: \eta(\tau_i) \eta(\tau_j') e^{-i\Omega(\tau_i - \tau_j')} W_{ji} (\Mtx_i(\tau_i), \Mtx_j'(\tau_j')) \vphantom{\infint }.
\end{align}
\end{widetext} 
In the weak coupling limit $(\lambda\ll 1)$, $P_E$ is the conditional transition (excitation) probability of the detector. Compared with the classical trajectory case, the new feature of this result is that $P_E$ contains Wightman function evaluated both locally along the individual trajectories ($i=j$ and henceforth, local terms), as well as nonlocally between any given pair of trajectories ($i\neq j$ and henceforth, nonlocal terms).\footnote{The nonlocal terms in the transition probably are identical to the ``nonlocal correlation'' terms in the reduced density matrix of two detectors, each interacting locally with the field via the Unruh-deWitt Hamiltonian along the individual classical paths $\Mtx_i(\tau_i)$ and $\Mtx_j(\tau_j)$. More specifically, the reduced bipartite density matrix is given by 
\begin{align}\label{reduceddensity}
    \hat{\rho}_{AB} &= \begin{pmatrix} 1 - P_A - P_B & 0 & 0 & M \\ 0 & P_B & L_{AB} & 0 \\ 0 & L_{AB}^\star  &  P_A & 0 \\ M^\star & 0 & 0 & 0  \end{pmatrix} + \mathcal{O}(\lambda^4) .
\end{align}
Here, $P_{A(B)}$ is the transition probability of detector $A(B)$, while both $M$ and $L_{AB}$ quantify different kinds of correlations between the two detectors. It is the $L_{AB}$ terms here that are equal to the nonlocal terms in the quantum-controlled Unruh-deWitt detector's transition rate. Meanwhile the $M$ term, which contains a time-ordered integral (i.e.\ from the second-order perturbative expansion) is usually interpreted as an amplitude for virtual particle exchange between the detectors.} 

\subsection{Wightman functions}
The behaviour of the Wightman functions $ W ^{ij}(\tau,\tau')$ has been discussed widely in the literature \cite{louko2006often,satz2007then,schlicht2004considerations,birrell1984quantum}. It was shown by Schlicht that the typical $i\varepsilon$-regularisation of the mode sum expansion of the fields, given by \cite{birrell1984quantum}
\begin{align}\label{ie}
    W ^{ij}(\tau,\tau') &= \lim_{\varepsilon\to 0} \frac{-1/4\pi^2}{({t}_i - {t}_j'-i\varepsilon)^2 - |\mathbf{x}_i - \mathbf{x}_j' |^2}
\end{align}
can lead to Lorentz non-invariant transition rates when the switching functions possess sharp cut-offs \cite{schlicht2004considerations}. Schlicht's solution was to spatially smear the field operator with a Lorentzian function -- which models a point-like detector in the limit $\varepsilon\to 0$ after integration over $(\tau,\tau')$ -- yielding the regularised result,
\begin{align}\label{schlicht}
   W ^{ij}(\tau,\tau') &= \frac{1/4\pi^2}{\big(\Mtx_i - \Mtx_j' - i \varepsilon( \dot{\Mtx}_i + \dot{\Mtx}_j') \big)^2}.
\end{align}
where $\dot{\Mtx}_i'$ is the 4-velocity of the detector evaluated at $\tau'$ along the $i$th trajectory of the superposition. In Sec.\ \ref{Superpose}, we utilise the Lorentzian-smeared regularisation, Eq.\ (\ref{schlicht}), to derive the conditional excitation probabilities and instantaneous transition rates of a detector travelling in a two-trajectory superposition. While \textit{regulator-free} expressions for the instantaneous transition rate have been derived previously \cite{louko2006often,satz2007then}, for the present work, Eq.\ (\ref{schlicht}) is convenient for obtaining numerical results. 

\subsection{Detector Transition Rates}
The time-dependent behaviour of the detector will be nontrivially affected by field correlations between different trajectories in the superposition. This motivates us to study the instantaneous transition rate of the detector while the interaction is still on. Rather than taking the $\tau \to \infty$ limit in Eq.\ (\ref{rhod}), which models the measurement as occurring in the asymptotic future, we consider the evolution of the system up to the time $\tau_F$, so that the transition probability from Eq.\ (\ref{excitation}) takes the form,
\begin{align}\label{NEEE}
 P_E &= \frac{\lambda^2}{N^2} \sum_{i=1}^N \sum_{j=1}^N \int_{-\infty}^{\tau_F} \D \tau_i \int_{-\infty }^{\tau_F} \eta(\tau_i) \eta(\tau_j') e^{-i\Omega(\tau_i - \tau_j')} W_{ji}(\tau_i,\tau_j'),
\end{align}
where we have made the simplification $W_{ji}(\Mtx_i(\tau_i), \Mtx_j'(\tau_j')) \equiv W_{ji}(\tau_i, \tau_j')$. This amounts to introducing a cut-off in the switching function at the proper time $\tau_F$, which can be understood as a strong detector-observer interaction. Differentiating Eq.\ (\ref{NEEE}) with respect to $\tau_F$, using the identity $W_{ji}(\tau_i, \tau_j' ) W_{ij}^\star(\tau_j',\tau_i)$ and making a change of variables $s = \tau_i - \tau_j'$ yields the following expression,
\begin{widetext} 
\begin{align}\label{switch}
    \dot{P}_E &= 2 \mathrm{Re} \left[ \frac{\lambda^2}{N^2} \sum_{i=1}^N \sum_{j=1}^N \eta(\tau_F) \int_0^\infty \D s \:e^{-i\Omega s} \eta(\tau_F - s)W_{ji}(\tau_F,\tau_F - s)  \right] ,
\end{align}
where we have denoted $\dot{P}_E = \D P_E/\D \tau$. We choose $\eta(\tau_i)$ to be a Gaussian switching function, $\eta(\tau_i) = e^{-\tau_i^2/2\sigma^2}$ where $\sigma$ is a characteristic timescale for the interaction. In the infinite interaction time limit, $\sigma\to \infty$, the transition rate reduces to
\begin{align}\label{past}
    \dot{P}_E &= 2 \mathrm{Re} \left[ \frac{\lambda^2}{N^2} \sum_{i=1}^N \sum_{j=1}^N \int_0^\infty \D s \:e^{-i\Omega s}W_{ji}(\tau_F,\tau_F - s) \right] .
\end{align}
As with $P_E$, the instantaneous transition rate $\dot{P}_E$ contains local ($i=j$) and nonlocal ($i\neq j$) terms.

It has been noted previously that Eq.\ (\ref{NEEE}) may be interpreted as the fraction of identically prepared detectors within a single ensemble that have undergone a transition after observation at time $\tau_F$ \cite{louko2006often,langlois2006causal,louko2008transition}. Since any observation alters the state of the system, Eq.\ (\ref{NEEE}) no longer carries this interpretation after the measurement. Therefore, Eq.\ (\ref{past}) compares the fraction of excited detectors in one ensemble, measured at $\tau_F + \delta \tau_F$, with that of another identically prepared ensemble measured at $\tau_F$, in the limit $\delta \tau_F \to 0^+$. A characteristic of Eq.\ (\ref{past}) is that it may be negative for certain values of $\tau$ \cite{louko2006often,langlois2006causal,louko2008transition}. To understand this, consider the final (unnormalised) detector-field state after a conditional measurement of the control in the state $|\mu \rangle$, given by
\begin{align}\label{25}
    |\psi\rangle_\mathrm{FD} &= \underbrace{ \bigg( \mathds{I} - \frac{\lambda^2}{N} \sum_{i=1}^N \infint\D \tau_i \: \eta(\tau_i)e^{i\Omega\tau_i} \int_{-\infty}^{\tau_i}\D \tau_i' \:  \eta(\tau_i') e^{i\Omega\tau_i'} \hat{\Phi} (\Mtx_i(\tau_i)) \hat{\Phi}(\Mtx_i'(\tau_i'))  \bigg)}_{\hat{\alpha}} \ranglem |g\rangle  \non 
    \\
    & - \underbrace{ \frac{i\lambda}{N} \sum_{i=1}^N \infint \D \tau_i \: \eta(\tau_i) e^{-i\Omega\tau_i}\hat{\Phi}(\Mtx_i(\tau_i) ) }_{\hat{\beta}} \ranglem |e\rangle 
\end{align}
\end{widetext} 
where $\hat{\alpha}\ranglem$ and $\hat{\beta}\ranglem$ are orthogonal field states. Given a measurement of the system at the proper time $\tau_F$, the detector is in a superposition of its ground and excited states, weighted by the amplitudes $\alpha$ and $\beta$. Importantly, these amplitudes need not be monotonic with $\tau_F$, as they contain field operators pulled back to the different trajectories. In spacetime regions where the dynamics change rapidly, the interplay between these terms may induce destructive interference that decreases the probability of excitation. In this way, the instantaneous transition rate in Eq.\ (\ref{past}) may take on negative values, whereas Eq.\ (\ref{NEEE}), representing the transition probability, is strictly positive. Moreover, since transitions in the detector arise from products of first-order terms in the perturbative expansion of $\hat{U}$, negative transition rates cannot be interpreted as arising from second-order processes such as detector excitation followed by emission. 

In Sec.\ \ref{Superpose}, we evaluate Eq.\ (\ref{past}) numerically to obtain the instantaneous transition rate for the quantum-controlled detector. Our analysis of the conditional excitation probability, obtained for a short interaction with a Gaussian switching function, is qualitatively different to that of the instantaneous transition rate, which is analysed in the infinite interaction time limit and measured while the interaction is still on. 

\section{Superpositions of Rindler Trajectories}\label{Superpose}

\begin{figure*}
    \centering
    \includegraphics[width=0.725\linewidth]{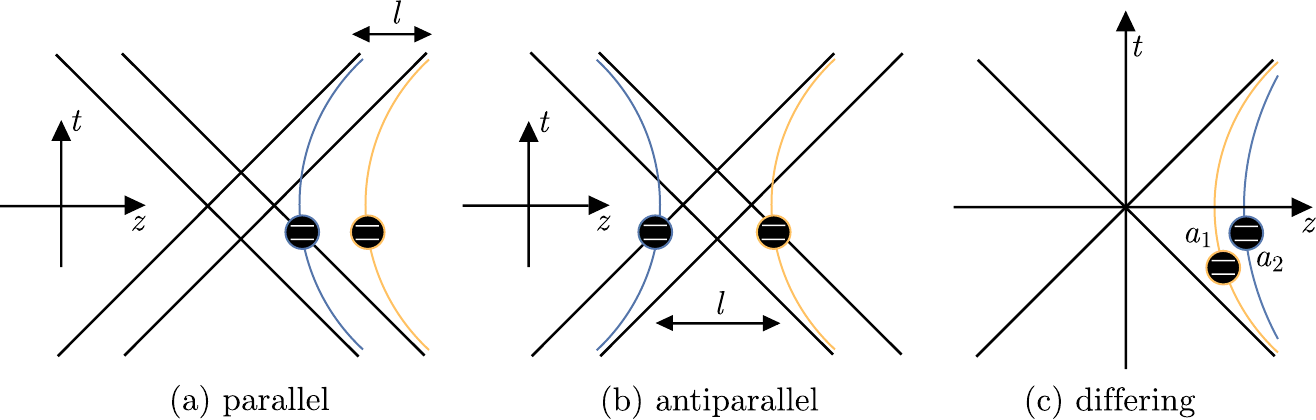}
    \caption{Schematic diagrams of the (a) parallel, (b) antiparallel, and (c) differing acceleration trajectory superpositions.}
    \label{fig:schematictraj}
\end{figure*}

As an application of our general, perturbative expressions, let us consider a UdW detector travelling in a superposition of two uniformly accelerated trajectories, in three classes of trajectory configurations. The coordinates parametrising ``parallel accelerations'' are given by 
\begin{align}\label{eq8.22}
    z_1 &= \frac{1}{a} \left( \cosh ( a \tau ) - 1 \right) + l / 2,
    \\
    z_2 &= \frac{1}{a} \left( \cosh (a \tau) - 1 \right) - l/2 ,
    \\
    t_1 = t_2 &= \frac{1}{a} \sinh (a \tau) ,
\end{align}
while for ``antiparallel accelerations'', we have 
\begin{align}
    z_1 &= \frac{1}{a} \left( \cosh (a \tau) - 1 \right) + l / 2 ,
    \\
    z_2 &= - \frac{1}{a} \left( \cosh (a \tau ) - 1 \right) - l/ 2,
    \\
\label{eq8.27}
    t_1 = t_2 &= \frac{1}{a} \sinh (a \tau ) .
\end{align}
In Eq.\ (\ref{eq8.22})--(\ref{eq8.27}), $a$ is the proper acceleration of the detector, $\tau$ its proper time, while $l$ defines the distance of closest approach as measured by an inertial observer along an arbitrary trajectory with constant $z$ (spacetime diagrams shown in Fig.\ \ref{fig:schematictraj}) \cite{salton2015acceleration}. The third class of configurations is a superposition of trajectories with differing accelerations and sharing a common Rindler horizon (spacetime diagrams shown in Fig.\ \ref{fig:schematictraj}),
\begin{align}\label{eqn:29}
    z_1 &= \frac{1}{a_1} \cosh (a_1 \tau) ,
    \vt 
    \\
    z_2 &= \frac{1}{a_2} \cosh(a_2 \tau) ,
    \vt 
    \\ \label{eqn:30}
    t_1 &= \frac{1}{a_1} \sinh(a_1\tau),
    \vt 
    \\
    t_2 &= \frac{1}{a_2} \sinh ( a_2\tau).
\end{align} 
Here, $a_1 \neq a_2$ are the superposed proper accelerations of the detector. In all cases, the other spatial coordinates are taken to be constant, and thus do not contribute to the dynamics. 

\subsection{Transition Probabilities: Parallel and Antiparallel Trajectories}
Let us consider first the detector transition probability for a superposition of parallel and antiparallel trajectories. We can evaluate the transition probability with respect to the proper time on the respective trajectories, allowing us to drop the subscripts on each trajectory:
\begin{align}
    P_E &= \frac{\lambda^2}{4} \sum_{i,j=1}^{N=2} \infint\D \tau \D \tau' \: \eta(\tau) \eta(\tau') e^{-i\Omega(\tau - \tau')} W_{ij}(\tau, \tau')  .
\end{align}
\begin{widetext}
The Wightman functions for the parallel trajectory superposition are given by\footnote{To obtain Eq.\ (\ref{eq8.35}) to (\ref{eq8.39}), we have utilised the identities:
\begin{align}
    \cosh (x) - \cosh(y) &= 2 \sinh \left(\frac{x-y}{2} \right) \sinh \left( \frac{x+y}{2} \right) ,
    \\
    \cosh(x) + \cosh (y) &= 2 \cosh \left( \frac{x-y}{2} \right) \cosh \left( \frac{x + y}{2} \right) ,
    \\
    \sinh(x) -  \sinh(y) &= 2 \sinh \left(\frac{x-y}{2} \right) \cosh \left(\frac{x+y}{2} \right) ,
    \\
    \sinh(x) + \sinh(y) &= 2 \cosh\left( \frac{x-y}{2} \right) \sinh\left( \frac{x+y}{2} \right) .
\end{align}
}
\begin{align}\label{eq8.35}
    W_{11}(s ) = W_{22}(s )  &= - \frac{1}{16\pi^2} \frac{1}{\beta(s)^2}
    \vphantom{\bigg]^{-1}},
    \\
    W_{12}(s,p) &= - \frac{1}{4\pi^2} \bigg[ 4 \cosh^2(ap/2) \beta(s)^2 - \Big( 2 \sinh (ap/2) \beta(s) + l \Big)^2 \bigg]^{-1},
    \\
    W_{21}(s,p) &= - \frac{1}{4\pi^2} \bigg[ 4 \cosh^2(ap/2) \beta(s)^2 - \Big( 2 \sinh (ap/2) \beta(s) - l \Big)^2 \bigg]^{-1},
\end{align}
while for the antiparallel trajectory superposition, we have 
\begin{align}
    W_{11}(s) = W_{22}(s) &= - \frac{1}{16\pi^2} \frac{1}{\beta(s)^2 }
    \vphantom{\bigg]^{-1}} ,
    \\
    \label{eq8.39}
    W_{12}(s,p) = W_{21}(s,p) &= - \frac{1}{4\pi^2} \bigg[ 4 \cosh^2 ( ap/2) \beta(s)^2 - \Big( 2\cosh (ap/2) \alpha(s) - 2a^{-1} + l \Big)^2 \bigg]^{-1} .
    \vphantom{\bigg]^{-1}}
\end{align}
\end{widetext} 
In the above, we have defined 
\begin{align}
    \beta(s) &= a^{-1}\sinh(as/2) - i \varepsilon \cosh(as/2),
    \vt 
    \\
    \alpha(s) &= a^{-1} \cosh(as/2) - i \varepsilon \sinh(as/2)
    \vt ,
\end{align}
where $s = \tau - \tau'$ and $p = \tau + \tau'$. To compute the transition probability, we consider the simple case of a Gaussian switching function, $\eta(\tau) = \exp (- \tau^2/2\sigma^2) $ with $\sigma\ll a^{-1}$ i.e.\ a short detector-field interaction. The terms $P_{ij,E}$ (where $i,j = 1, 2$) in the excitation probability become,
\begin{align}
    P_{ij,E} &= \xi_0 \infint\D p \D z \:  e^{-\frac{p^2}{4\sigma}} e^{-\frac{z^2}{4\sigma^2}} W_{ij}(z - 2 i \Omega \sigma^2,p ) + \mathrm{res.}
\end{align}
having converted the complex Gaussian into a real Gaussian by shifting the axis of integration and we have defined $\xi_0 = \lambda^2e^{-\Omega^2\sigma^2}/8$. We have included an additional residue term that arises when this shifting crosses a poles in the complex plane.\footnote{Below, we generally focus, for simplicity, on regimes in which the residue terms vanish.} Now, in the saddle-point approximation $(\sigma \ll a^{-1})$, the switching function Gaussians are sharply peaked around $ p = s = 0$. This allows us to treat these switching functions as approximating Dirac--$\delta$ functions, yielding the simplified expression:
\begin{align}
    P_{ij,E} &\simeq \frac{1}{2}\lambda^2 e^{-\Omega^2\sigma^2} \pi \sigma^2 W_{ij}(-2i\Omega\sigma^2, 0 ) + \mathrm{res.}
    \vphantom{\bigg]^{-1}}
\end{align}
where $\Omega \sigma$ is assumed to be small. Now, noting that shifting the contour in the complex plane does not cross poles in the Wightman function if we restrict our analysis to $a\Omega \sigma^2 < \pi$ (see Sec.\ \ref{ch8appendix}), we find the following expressions for the superposition of parallel accelerations,
\begin{widetext} 
\begin{align}
\label{eq8.48}
    P_{11,E} = P_{22,E} &= \frac{\lambda^2 a^2 \sigma^2 e^{-\Omega^2 \sigma^2}}{32\pi} \mathrm{csc}^2(a\Omega \sigma^2 ) ,
    \\
    P_{12,E} = P_{21,E} &= \frac{\lambda^2 a^2 \sigma^2 e^{-\Omega^2\sigma^2}}{32\pi} \bigg[ \sin^2 (a \Omega \sigma^2 ) + (al/2 ) \bigg]^{-1} ,
\end{align}
while for the superposition of antiparallel accelerations, 
\begin{align}
\label{eq8.50}
    P_{11,E} = P_{22,E} &= \frac{\lambda^2 a^2 \sigma^2 e^{-\Omega^2\sigma^2}}{32\pi} \mathrm{csc}^2(a\Omega\sigma^2 ) ,
    \\
    P_{12,E} = P_{21,E} &= \frac{\lambda^2a^2\sigma^2 e^{-\Omega^2\sigma^2}}{32\pi} \bigg[ \sin^2(a\Omega \sigma^2 ) + \Big( \cos ( a \Omega \sigma^2 )  -1 + (al/2) \Big)^2 \bigg]^{-1} .
\end{align}
Altogether, the transition probability of the parallel trajectory superposition is given by 
\begin{align}
    P_E &= \frac{\lambda^2a^2\sigma^2 e^{-\Omega^2\sigma^2}}{16\pi} \bigg[ \mathrm{csc}^2(a\Omega \sigma^2 ) + \frac{1}{\sin^2 ( a \Omega \sigma^2 ) + (al/2)^2} \bigg],
\end{align}
while for the antiparallel case, we have
\begin{align}
    P_E &= \frac{\lambda^2 a^2 \sigma^2 e^{-\Omega^2\sigma^2}}{16\pi} \bigg[ \mathrm{csc}^2(a\Omega\sigma^2) + \Big( \sin^2( a\Omega\sigma^2) + \Big( \cos (a\Omega\sigma^2) - 1 + (al/2) \Big)^2 \Big)^{-1} \bigg] .
\end{align}
\end{widetext} 
Let us examine some limits of these results. For the parallel superposition, when $l = 0$, the transition probability trivially reduces to that of a single detector on an accelerated trajectory with acceleration $a$:
\begin{align}
    P_E &= \frac{\lambda^2a^2 \sigma^2 e^{-\Omega^2\sigma^2}}{8\pi} \mathrm{csc}^2(a\Omega\sigma^2) .
\end{align}

\begin{figure*}
    \centering
    \includegraphics[width=0.75\linewidth]{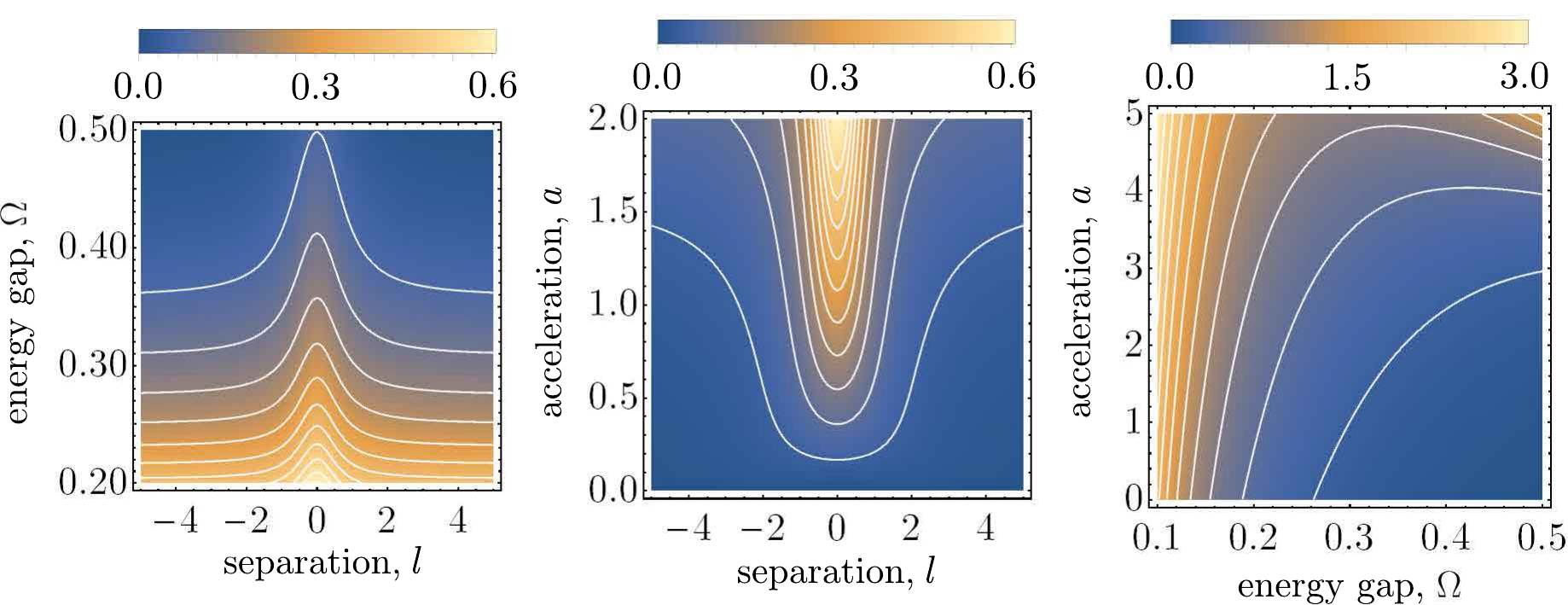}
    \caption[Transition probabilities for the detector in a superposition of parallel accelerations]{Transition probability, $P_E/\lambda^2$, for the detector in a superposition of parallel accelerations. We have used the parameters (left) $a\sigma = 2$, (middle) $\Omega\sigma = 0.2$, and (right) $l/\sigma = 0.1$ respectively. }
    \label{fig:paralleltp}
\end{figure*}
\begin{figure*}
    \centering
    \includegraphics[width=0.75\linewidth]{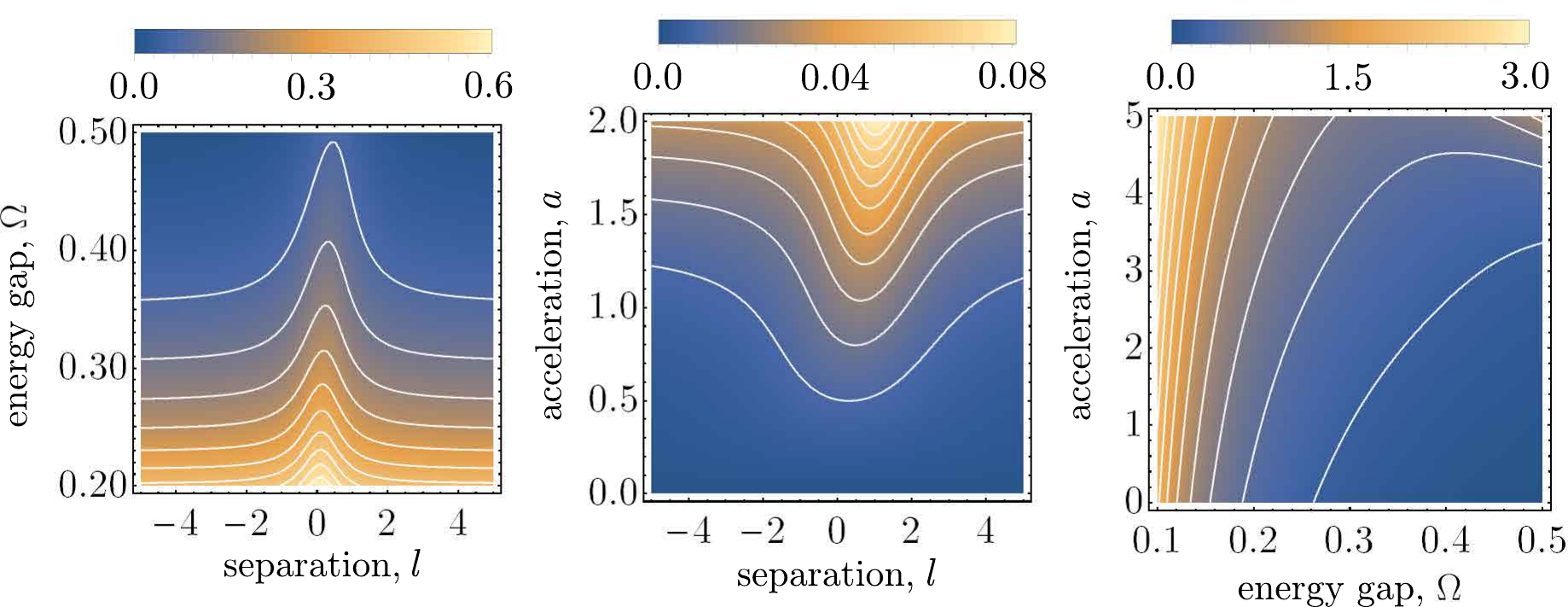}
    \caption[Transition probabilities for the detector in a superposition of antiparallel accelerations]{Transition probabilities, $P_E/\lambda^2$, for the detector in a superposition of antiparallel accelerations. We have used the parameters (left) $a\sigma = 2$, (middle) $\Omega\sigma = 0.8$, and (right) $l/\sigma = 0.1$ respectively.}
    \label{fig:antiparalleltp}
\end{figure*}

Meanwhile, taking $l \to \infty$ causes the interference term to vanish:
\begin{align}
    P_E &= \frac{\lambda^2 a^2\sigma^2 e^{-\Omega^2\sigma^2}}{16\pi} \mathrm{csc}^2(a\Omega\sigma^2),
\end{align}
yielding half of the value for a single detector on a classical trajectory. At finite separations, the field correlations between the trajectories inhibit detector excitations relative to the single trajectory case. This result may have implications for entanglement harvesting as we note in Sec.\ \ref{conclusion}. Generally, the amount of entanglement that can be extracted from quantum fields onto bipartite UdW detector systems is limited by the local noise surrounding each detector, and thus, the inhibition of excitations in the detector should allow two superposed detectors to become more strongly entangled, as recent results have suggested \cite{henderson2020quantum}.

In the case of the antiparallel superposition, the $l = 0$ limit yields
\begin{align}
    P_E &= \frac{\lambda^2 a^2\sigma^2 e^{-\Omega^2\sigma^2}}{16\pi} \bigg[ \mathrm{csc}^2(a\Omega\sigma^2) + \frac{1}{4}\mathrm{csc}^2(a\Omega\sigma^2/2) \bigg].
\end{align}
Unlike the parallel case, there is nothing geometrically unique about the $l = 0$ limit, due to the way that the trajectories are configured. On the other hand, when $l \to \infty$, the nonlocal interference term vanishes also in this case, leaving
\begin{align}
    P_E &= \frac{\lambda^2a^2\sigma^2 e^{-\Omega^2\sigma^2}}{16\pi} \mathrm{csc}^2( a \Omega \sigma^2 ) ,
\end{align}
which is half of the transition probability of a single detector. Finally, it is worth noting that the contributions to the transition probability along either of the paths in superposition, Eq.\ (\ref{eq8.48}) and (\ref{eq8.50}), are characteristically nonthermal. This is because the short detector-field interaction will generally not allow the detector to reach thermal equilibrium. 

In Fig.\ \ref{fig:paralleltp}, we have plotted the transition probability of the detector in the parallel configuration. The left and middle subplots illustrate how in general, excitations in the detector are suppressed as the separation between the superposed trajectories increases. This may be interpreted as a kind of destructive interference between the trajectories. As one should expect, the transition probability decays with the energy gap, $\Omega$, while increasing with larger accelerations. The main difference between the parallel and antiparallel configurations (the latter plotted in Fig.\ \ref{fig:antiparalleltp}) is the asymmetry in $l$ for the latter. Indeed, for certain regions of the $(\Omega, a)$ parameter space, the choice of $l < 0$ (geometrically, when the antiparallel trajectories overlap for some region in spacetime near the origin of coordinates) significantly inhibits excitations in the detector, compared with a choice of $l > 0$ with the same magnitude, which does not do so to the same degree. Evidently, the physical configuration of the superposed trajectories traversed by the detector has a significant effect on its response to the field.

\subsection{Transition Rates: Parallel and Antiparallel Trajectories}
\begin{figure*}
    \centering
    \includegraphics[width=0.65\linewidth]{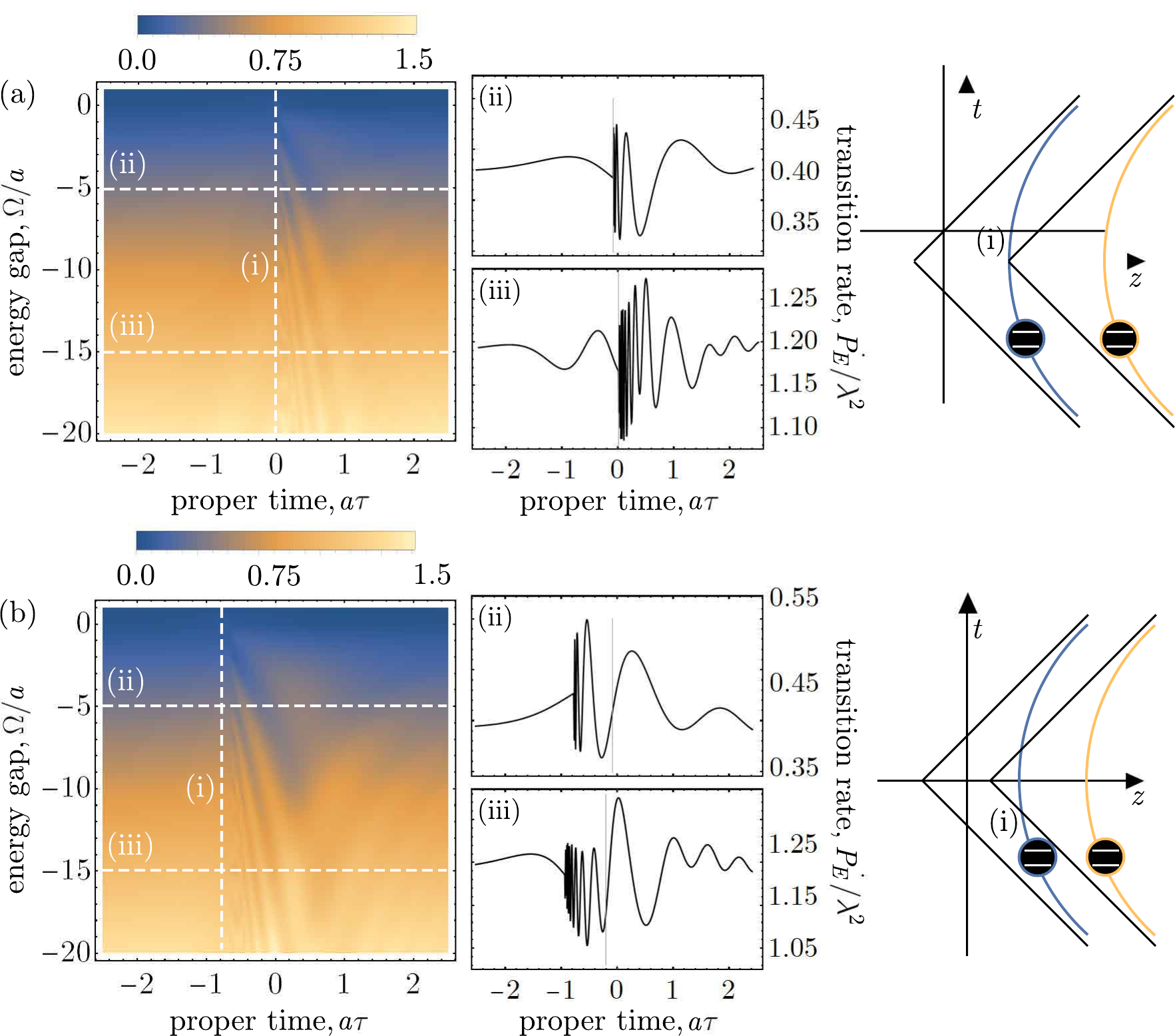}
    \caption[Transition rate of the detector in a superposition of parallel accelerations]{Transition rate of the detector in a superposition of parallel accelerations, for (a) $l / a = 1.0$, and (b) $l / a = 0.5$. In both plots (i) indicates the horizon-crossing point of the trajectories, (ii) as a $\Omega$-slice corresponding to $\Omega = -5$, and as is (iii) for $\Omega = -15$. The schematic spacetime diagrams illustrate the horizon-crossing point for the respective setups.}
    \label{fig:parallelrate}
\end{figure*}

Next, we are interested in the instantaneous transition rate of the detector as it travels along the superposition of accelerated trajectories. This rate is given by
\begin{align}\label{transitionrate}
    \dot{P}_E &= \frac{1}{2} \mathrm{Re} \int_0^\infty \D s \:e^{-i\Omega s} \Big(W_{11}(s) + W_{22}(s) 
    \non 
    \\
    & \qquad + W_{12} (\tau, \tau - s) + W_{21}(\tau, \tau - s ) \Big) .
\end{align}
To plot the transition rate, we utilise a numerical analysis to evaluate the integrals in Eq.\ (\ref{transitionrate}). The instantaneous transition rate of the detector in a superposition of parallel accelerations is shown in Fig.\ \ref{fig:parallelrate}. 

\begin{figure*}
    \centering
    \includegraphics[width=0.65\linewidth]{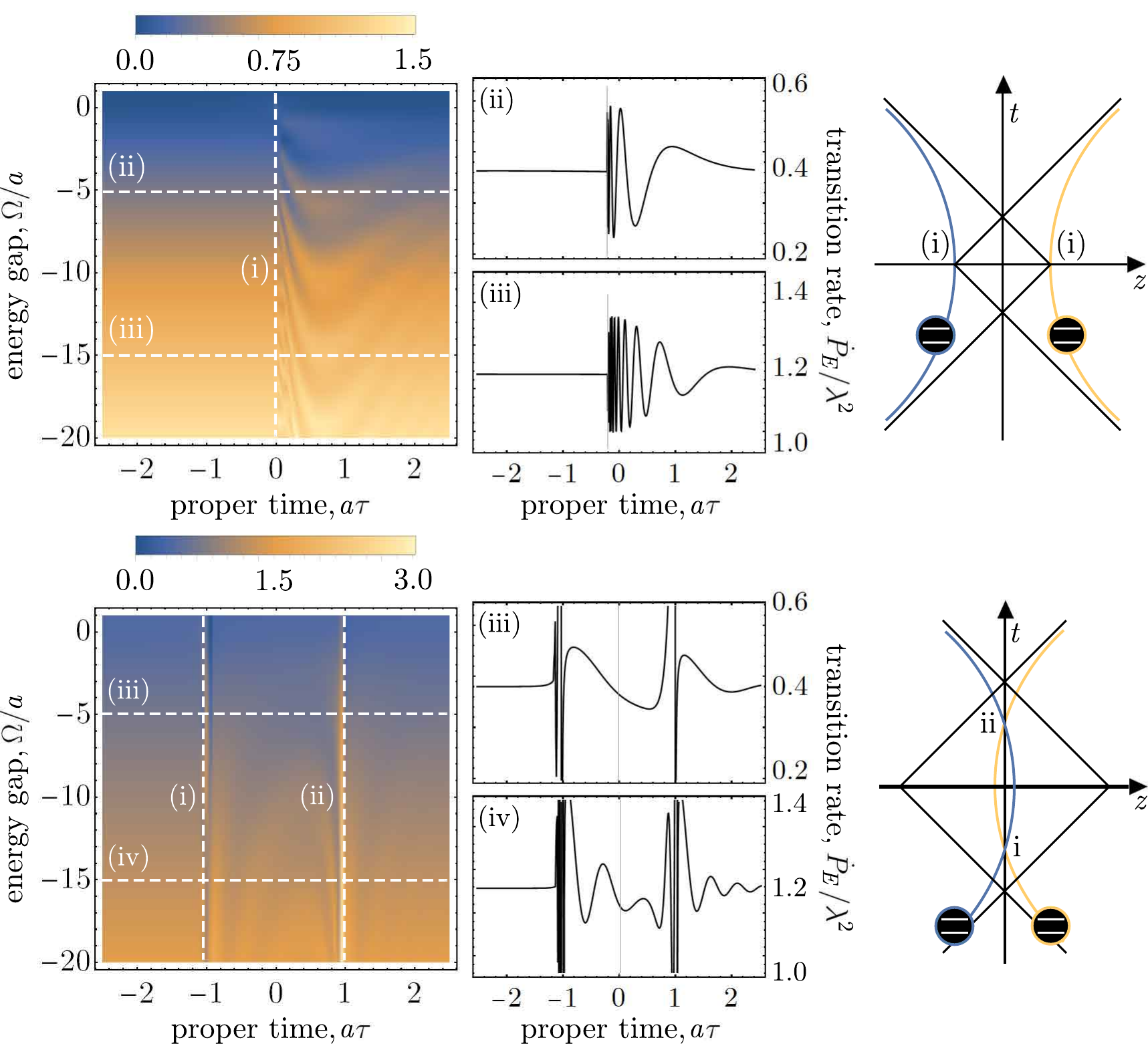}
    \caption[Transition rate of the detector in a superposition of antiparallel accelerations]{Transition rate of the detector in a superposition of antiparallel accelerations. For the two subplots. we have chosen (a) $l /a = 1.0$ and (b) $l / a = -1.0$. The labels (ii) and (iii) denote $\Omega$-slices at $\Omega = -5$ and $\Omega = -15$ respetively. In (a), the label (i) denotes the horizon-crossing point of the two trajectories, while in (b), (i) and (ii) denote the points at which the trajectories cross. }
    \label{fig:antiparallelrate}
\end{figure*}

We have primarily displayed the transition rate behaviour for $\Omega<0$, for which $\dot{P}_E$ represents the rate of stimulated emission given that the detector is prepared in its excited state. Several key observations can be made. First, the energy spectrum of the detector in Fig.\ \ref{fig:parallelrate} is not just the Planckian distribution of a single uniformly accelerated detector, 
\begin{align}
    \dot{P}_E &\propto \frac{\Omega}{2\pi} \frac{1}{\exp(2\pi \Omega/a) - 1 },
\end{align}
but exhibits time-dependent behaviour that also depends on the energy gap, $\Omega$. The transition rate experiences a sudden onset of oscillations at a certain proper time. These dynamics begin to manifest as the left-most trajectory ($t_1,z_1$) crosses the light-like extension of the Rindler horizon of the right-most trajectory $(t_2,z_2)$. This rapid behaviour does not reappear as the detector recedes from the origin--that is, as the right-most trajectory passes through the lightcone of the left-most trajectory. We conjecture that it is produced by the build-up of detector-field interactions that are transmitted through the field from the asymptotic past, i.e.\ at the origin of the right-most trajectory. For large $\tau$, the transition rate equilibrates towards the stationary value of a single detector.

The novel behaviour of the transition rate, most prominently observed at the horizon-crossing event, arises because the causal relationship between the trajectories is asymmetric in time. That is, the causal influence of the right-most trajectory with the left-most trajectory, mediated by the nonlocal field correlations between them, is not identical to that of the left-most trajectory upon the right-most trajectory. A comparable scenario without this asymmetry is a detector superposed along two inertial trajectories separated by a fixed distance $l$ and immersed in a thermal bath at the finite temperature $T_T = a(2\pi)^{-1}$. The causal relationship between the trajectories is now symmetric, and does not possess the Rindler wedge structure of the accelerated trajectories. While the response of the detector along either inertial trajectory is identical to that for the uniformly accelerated detector in the Minkowski vacuum, the nonlocal terms take the form \cite{birrell1984quantum,weldon2000thermal}
\begin{align}\label{eq8.66}
    W_T(s) &= \frac{a}{16\pi^2 l} \Big(  \coth \left( \frac{a}{2}(l-s + i \varepsilon )\right) 
    \non 
    \\
    & \qquad + \coth \left( \frac{a}{2} (l+s - i \varepsilon ) \right) \Big).
\end{align}
Notably, $W_T(s)$ is time-translation invariant (depending only on $s= \tau - \tau')$, implying that the transition rate of such a detector is likewise time-independent. This demonstrates that the physical trajectories of the detector nontrivially affect its response by changing the detector’s proper time, along with the causal relations and ordering of the interactions along the individual paths.

The instantaneous transition rate of the detector in a superposition of antiparallel accelerations is shown in Fig.\ \ref{fig:antiparallelrate}, for (a) $0 < l < 2a^{-1}$ and (b) $l < 0$. In (a), the trajectories are not spacelike separated and do not overlap spatially. In a comparable manner to the parallel motion, the transition rate is approximately time-independent until the field perturbation produced by the detector interactions from the asymptotic past becomes causally connected with the spacetime region of the other trajectory, at which the transition rate becomes highly oscillatory. In (b), the superposed trajectories do overlap. We observe interesting behaviour near these crossing points, whereby the detector transition rate oscillates rapidly. This reflects the highly nontrivial properties of the field correlations at short ranges, which are accessible to the detector through its nonlocal coupling to the field along its superposed trajectories. Finally, for $l>2a^{-1}$ (not shown), the trajectories are spacelike separated and the effect of the nonlocal terms in $\dot{P}_E$ decreases with $l$. In this scenario, the Rindler horizon of either trajectory never intersects the other, suppressing the dramatic dynamical effects observed at the horizon-crossing event for the causally connected trajectories.  
\subsection{Detector Thermalisation}
It is interesting to consider whether the detector ever exhibits an exactly thermal response, as is the case for a single trajectory. The transition rate must satisfy the detailed balance form of the Kubo-Martin-Scwhinger (KMS) condition \cite{haag1967equilibrium}, 
\begin{align}\label{kubo}
    \frac{\dot{P}_E(\Omega)}{\dot{P}_E(-\Omega)} &= \exp ( - 2\pi \Omega/ a ) .
\end{align}
We find that for both the parallel and antiparallel configurations, the KMS condition is only satisfied in the limit $l\to \pm\infty$ (and this is even for superposed trajectories having the same magnitude of proper acceleration). In this regime, the field correlations between the individual trajectories is zero, characterised by the vanishing of the nonlocal Wightman functions in Eq.\ (\ref{transitionrate}). The transition rate in both cases reduces to
\begin{align}
    \dot{P}_E &=  \mathrm{Re}\int_0^\infty \D s \:e^{-i\Omega s} W_{11}(s) = \frac{\Omega}{4\pi} \frac{1}{\exp(2\pi\Omega/a)-1},
\end{align}
which is half of the transition rate for a single uniformly accelerated detector, in agreement with the results found for the excitation probability. Importantly, for finitely separated trajectories in the superposition, the presence of nonzero, time-dependent field correlations between them always alters the detector's response from being exactly thermal. This nonthermalisation is likely due to the inequivalent mode structure for the modes associated with the different Rindler wedges of the spatially translated trajectories.

\subsection{Transition Probabilities and Rates: Differing Accelerations}
Finally, we consider the detector travelling in a superposition of uniformly accelerated paths with differing proper accelerations $a_1$ and $a_2$, using the co-ordinates defined in Eq.\ (\ref{eqn:29}) and (\ref{eqn:30}). The nonlocal Wightman functions are given by 
\begin{widetext} 
\begin{align}
    W_{12}(\Mtx, \Mtx') &= - \frac{1}{4\pi^2} \bigg[ \Big( a_1^{-1} \sinh(a_1 \tau_1) - a_2^{-1} \sinh(a_2 \tau_2') - i\varepsilon \Big( \cosh(a_1 \tau_1) + \cosh(a_2\tau_2') \Big) \Big)^2
    \non 
    \\
    & - \Big( a_1^{-1} \cosh(a_1 \tau_1) - a_2^{-1} \cosh(a_2 \tau_2') - i\varepsilon \Big( \sinh(a_1 \tau_1) + \sinh(a_2 \tau_2') \Big) \Big)^2\bigg]^{-1} ,
    \\
    W_{21}(\Mtx, \Mtx') &= - \frac{1}{4\pi^2} \bigg[  \Big( a_2^{-1} \sinh(a_2 \tau_2) - a_1^{-1} \sinh(a_1 \tau_1') - i\varepsilon \Big( \cosh(a_2 \tau_2) + \cosh(a_1 \tau_1') \Big) \Big)^2 
    \non 
    \\
    & - \Big( a_2^{-1} \cosh(a_2 \tau_2) - a_1^{-1} \cosh(a_1 \tau_1') - i\varepsilon \Big( \sinh(a_2 \tau_2) + \sinh(a_1 \tau_1') \Big) \Big)^2 \bigg]^{-1} .
\end{align}
Notice in particular that time-translation invariance of the two Wightman function is broken, due to the different rates at which the proper time evolves along the two paths. For $a_1=a_2$, the Wightman functions reduce to that of a single accelerated trajectory. Unlike the parallel and antiparallel cases, shifting the contours of integration crosses poles in the Wightman functions. Nevertheless, the usual procedure can still be applied to obtain a semi-analytic result with the inclusion of residue contributions,
\begin{align}\label{diff}
    P_E &= \frac{\lambda^2 \sigma^2e^{-\Omega^2\sigma^2}}{32\pi}
    \left[ a_1^2 \mathrm{csc}^2(a_1 \sigma^2 \Omega ) + a_2^2 \mathrm{csc}^2(a_2 \sigma^2 \Omega) + \frac{8a_1^2a_2^2}{a_1^{2}+a_2^2 - 2a_1a_2\cos((a_1 + a_2) \sigma^2 \Omega)} \right] + \mathrm{res.}
\end{align}
\end{widetext} 
Since the residue contributions are difficult to obtain analytically, we leave a quantitative analysis of the excitation probability for future work. Nevertheless, by inspecting Eq.\ (\ref{diff}), we discover independent contributions to $P_E$ from the individual trajectories, as well as an interference term between them. It is also possible to analyse two simple cases. The first occurs in the limit when $a_1 = a_2$ whereby the residue contributions vanish, yielding the excitation probability for a single accelerated detector. The other occurs when one acceleration $(a_1 \to 0)$ approaches zero, for which the excitation probability is given by
\begin{align}
  P_E &= \frac{\lambda^2\sigma^2 e^{-\Omega^2\sigma^2}}{32\pi} \left[  \frac{1}{\Omega^2 \sigma^4}  + a_2^2 \mathrm{csc}^2(a_2\sigma^2\Omega) \right] 
    + \mathrm{res.}
\end{align}
In this form, the transition probability contains a constant contribution from the inertial trajectory, while the contribution from the acceleration is a quarter of that for a single accelerated detector. That the detector registers a nonzero particle count along the inertial trajectory can be understood as a consequence of the energy-time uncertainty relation, $\Delta t \Delta E \geq 1$. Since the detector-field interaction is highly localised in time, the contribution from the inertial trajectory is nonzero even when the field is in the vacuum state  \cite{menicucciPhysRevD.79.044027}. 

\begin{figure*}
    \centering
    \includegraphics[width=0.7\linewidth]{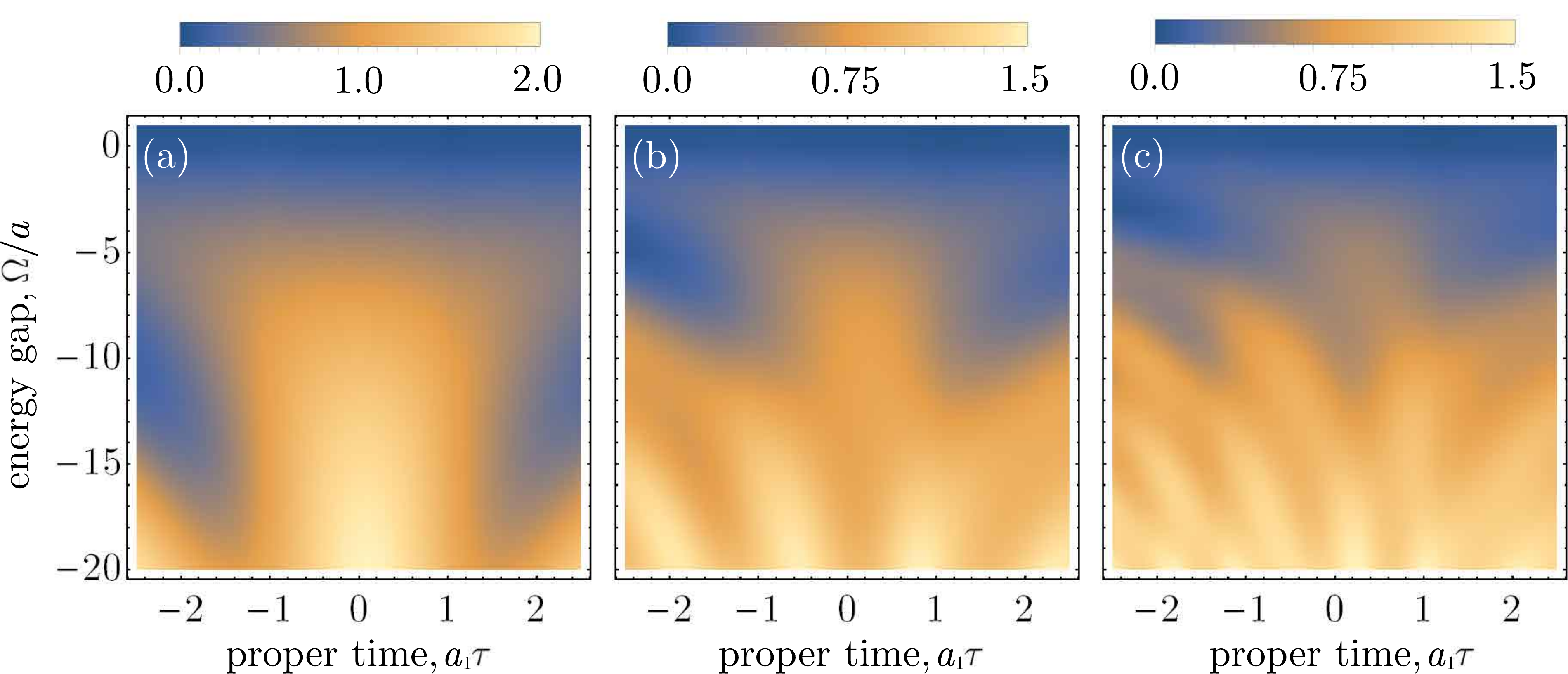}
    \caption[Transition rate of the detector in a superposition of differing accelerations]{Transition rate of the detector in a superposition of differing accelerations. The three subplots correspond to $a_1 = 1.0$ and (a) $a_2 = 0.9$, (b) $a_2 = 0.7$, and (c) $a_2 = 0.5$. }
    \label{fig:differingrate}
\end{figure*}

For the instantaneous transition rate, $\tau = \tau_1 = \tau_2$ represents the equal proper time along the respective trajectories at which the detector is measured, recalling Eq.\ (\ref{NEEE}). As shown in Fig.\ \ref{fig:differingrate}, the emission rate exhibits sustained oscillations since the distance between the paths of the superposition changes constantly according to an inertial observer. The frequency of these oscillations is blue-shifted (red-shifted) as the detector approaches (recedes from) the origin, indicating that the dynamics are affected by causal relationships between the trajectories, similar to those observed previously. 

\section{Coherence of the Superposition}\label{sec:general}
We previously considered the specific case where the control degree of freedom was measured in its initial state, $|\mu\rangle = (1/\sqrt{N})\sum_{i=1}^N | i \rangle$. However in general, the control need not be measured in the same superposition state in which it was initially prepared. For example, the detector-field interaction may cause the individual control states $|i\rangle$ associated with the respective trajectories to evolve by some relative phase. This is also analogous to the controllable phase shifts that commonly appear in Mach-Zehnder-type setups. Consider the general case wherein the control is measured in the final state
\begin{align}\label{controlfinal}
    |\nu \rangle &= \frac{1}{\sqrt{N}} \sum_{i=1}^N e^{-i\varphi_i}|i\rangle .
\end{align}
The final state of the detector is described by the following density matrix,
\begin{align}\label{rhodcontrol}
    \hat{\rho}_D &= \frac{1}{N^2} \sum_{i,j=1}^N e^{-i(\varphi_i - \varphi_{j}) } \hat{\rho}_{ij,E},
\end{align}
where the individual contributions $\rho_{ij,D}$ take the form of Eq.\ ( \ref {rhod}). For a superposition of two trajectories, the elements of the (unnormalised) density matrix are
\begin{align}\label{69}
    1 - P_G &= \frac{1}{2} \left( 1 + \cos(\delta \varphi ) \right) \left( 1 - \frac{P_{11,E} + P_{22,E}}{2} \right) ,
    \\ 
    \label{70}
    P_E &= \frac{1}{4} \left( P_{11,E} + P_{22,E} + 2 \cos(\delta\varphi) P_{12,E} \right) , 
\end{align}
where $\delta\varphi = \varphi_1 - \varphi_2$. Recall that the sum of Eq.\ ( \ref {69}) and ( \ref {70}) gives the normalisation of the final state of the detector, conditioned upon the control being measured in the superposition state, Eq.\ ( \ref {controlfinal}). Equivalently, this is the probability amplitude one obtains when measuring the control-detector-field state in the phase-shifted state Eq.\ ( \ref {controlfinal}):
\begin{align}
    \left\langle \nu \left| \mathrm{Tr}_{\Phi,D} \big[ \hat{\rho}_\mathrm{CFD} \big] \right| \nu \right\rangle &= 1 - P_G + P_E ,
\end{align}
where $\hat{\rho}_\mathrm{CFD} = | \mu \rangle\langle \mu |\otimes | 0_M \rangle\langle 0_M | \otimes | g \rangle\langle g |$ is the initial product state of the control, field, and detector degrees of freedom and $\mathrm{Tr}_{\Phi,D} [ \hdots ] $ denotes a trace over both the field and detector degrees of freedom. Explicitly, 
\begin{align}
    & \left\langle \nu \left| \mathrm{Tr}_{\Phi,D} \big[ \hat{\rho}_\mathrm{CFD} \big] \right| \nu \right\rangle 
    \non 
    \\
    & \:\:\: = \frac{1}{2} \left( 1 + \underbrace{ \left( 1 + P_{12,E} - \frac{P_{11,E} + P_{22,E}}{2} \right) }_V \cos(\delta \varphi) \right) .
\end{align}
In interferometric terms, the coefficient of the $\cos(\delta\phi)$ term, $V$, is the so-called visibility: the amplitude of oscillation as the relative phase $\delta\varphi$ is varied. Note here that in general, $V \lesssim \mathcal{O}(1)$, since the magnitude of the $P_{ij,E}$ terms is on the order of $\mathcal{O}(\lambda^2)$. Now, a nonzero visibility suggests that the control retains some coherence after the interaction, which in our case will be generally true. This further implies that the evolution of the system, mediated by the interaction, causes the control to become entangled with the detector-field system. When the local terms, $P_{11,E}$ and $P_{22,E}$, dominate over the interference terms, $P_{12,E}$, then the visibility will be reduced and decoherence of the control will result.  

\subsection{On the Control}
In Sec.\ \ref{udwmodel}, we introduced the control degree of freedom $i$ which initialises the superposition, without specific reference to a physical model. Several proposals for the physical implementation of UdW detector superpositions have been studied in the literature. For example, \cite{martin2011using} considers how a measurement of the Berry phase acquired by an accelerating detector may allow for a measurement of the Unruh effect at significantly lower accelerations than ordinarily required. In that paper, the authors propose an in-principle atomic interferometry setup \cite{bongs2019taking} which causes the detector to travel in a superposition of an inertial and accelerated path. In \cite{onuma2013mode,martin2013berry}, the authors consider another in-principle setup where single UdW detectors are sent through a beamsplitter which allows them to extract unknown information localised to cavities along the respective arms. \cite{martin2013berry} uses this to model a quantum thermometer, allowing the detector to measure the temperature of a cold reservoir relative to a hotter reservoir, without requiring that the detector reach thermal equilibrium. These works provide feasible schemes wherein our quantum-controlled detector model could be physically realised. 

Finally, we note that a detection of the interference effects presented in this paper would be on the same order of difficulty as a measurement of the Unruh effect itself. It is well-known that accelerations of $\sim 10^{20}$m/s$^2$ are required to register an Unruh temperature of $1$K -- hence, we expect that accelerations to this order or larger would be required to observe perturbations away from the expected detector response.   

\section{Conclusion}\label{conclusion}
In this paper, we have established a general framework for a UdW detector travelling in a quantum superposition of classical trajectories, by introducing an additional degree of freedom which could create and control such a superposition. To second-order in perturbation theory, we derived the final state of a detector traversing an arbitrary superposition of paths, and subsequently its conditional excitation probability and instantaneous transition rate. This final state, including the instantaneous transition rate, depends on \textit{local} two-point correlation functions evaluated along each individual trajectory, as well as \textit{non-local} or \textit{interference} terms between the trajectories. For particular scenarios involving two-trajectory superpositions of accelerated paths in parallel and anti-parallel motion, we derived semi-analytic expressions for the excitation probability of the detector.

Notably, we discovered that even for a superposition of paths with the same proper acceleration (parallel and anti-parallel trajectories), the final state of the detector differs from that of a single trajectory, and in particular is not thermal. In contrast, if the detector followed either of the individual trajectories, it would register the same thermal response. A thermal response is recovered for infinite separation between the trajectories in the superposition. Our numerical results for the instantaneous transition rate revealed novel interference effects not observed in the single detector scenario. In the parallel and anti-parallel scenarios, we discovered sudden periods of rapid, oscillatory behaviour in the transition rate, which revealed a dependence on causal dynamics between the trajectories. For the detector travelling in a superposition of proper accelerations, these causal relations induced Doppler-shifting of the oscillations in the emission rate as the detector approached, then receded from the origin. 

Our new approach can be directly applied to scenarios of fundamental interest in quantum field theory, cosmology and even quantum gravity. For example, it can be used to describe an UdW detector in a spacetime produced by a black hole in quantum superposition of different masses, or a detector in superposition of falling into a black hole and escaping it. Via the equivalence principle, we can also simulate spacetimes with entangled temporal order \cite{zych2019bell} by considering Rindler observers with entangled proper accelerations \cite{dimic2017simulating}. Other trajectories and spacetimes of interest, such as the de Sitter and anti-de Sitter geometries can also be studied by exploiting the well-known relationships between these spacetimes, Rindler geometry and conformal field theory \cite{birrell1984quantum,salton2015acceleration}.

From a broader perspective, our results suggest a connection between recent works studying coherent control of quantum channels \cite{oi2003interference,ebler2018enhanced,chiribella2019quantum,abbott2018communication,guerin2019communication}, relativistic quantum information \cite{mann2012relativistic}, and quantum thermodynamics \cite{vinjanampathy2016quantum}. Here, the interaction between an UdW detector and a quantum field, facilitated by the coherent control of the detectors' trajectory, directly results in a quantum control of different channels acting on the detector.  It has been shown \cite{chiribella2019quantum,abbott2018communication,guerin2019communication} that quantum control can result in increased channel capacities. Quantum-controlled UdW detectors can thus be exploited from the perspective of reducing the unavoidable noise experienced by non-inertial parties (due to the Unruh effect) in any relativistic quantum information setting. This expectation is further corroborated by recent findings in \cite{henderson2020quantum} showing that quantum control of the interaction time of inertial UdW detectors with a quantum field allows the detectors to become entangled in scenarios where this is otherwise impossible \cite{simidzija2018general}.

Secondly, from the perspective of quantum thermodynamics, the quantum-controlled UdW model introduces a new scenario, a quantum control of thermalisation channels. Quantum aspects of temperature are of high interest and relevance to this field \cite{miller2018energy} and our approach paves the way for answering foundational questions about the physical meaning and phenomenology associated with coherent control of temperatures. Already, the present results hint at a rich structure of the problem, as we have found that the quantum control of channels yielding the same Unruh temperature, in general do not result in any thermalisation of the system. 

Note added: towards the completion of this work we became aware of an independent study on a similar topic by Barbado, Castro-Ruiz, Apadula and Brukner, see arXiv:2003.12603.

\section{Acknowledgements}
The authors would like to thank Robert Mann for fruitful discussions. We acknowledge support from the Australian
Research Council Centre of Excellence for Quantum
Computation and Communication Technology (Project
No.\ CE170100012) and DECRA Grant DE180101443.

\appendix 
\begin{widetext}
\section{Complex integration}\label{ch8appendix}
In this appendix, we derive the pole-crossing constraint on the Wightman functions for the parallel and antiparallel superpositions, as well as the interferometric visibility of the control. 

\subsection{Pole Constraints: Parallel Accelerations}
In the parallel superposition, the first Wightman function takes the form, 
\begin{align}
    W_{12}(s,p) &= - \frac{1}{4\pi^2} \bigg[ 4 \cosh^2(ap/2) \beta(s)^2 - \Big( 2 \sinh (ap/2) \beta(s) + l \Big)^2 \bigg]^{-1},
\end{align}
where
\begin{align}
    \beta(s) &= a^{-1}\sinh(as/2) - i \varepsilon \cosh(as/2),
    \vt 
    \\
    \alpha(s) &= a^{-1} \cosh(as/2) - i \varepsilon \sinh(as/2)
    \vt .
\end{align}
This can be expressed in the following form:
\begin{align}
    W_{12}(s,p) &= - \frac{1}{4\pi^2} \bigg[ \Big( 2a^{-1} \sinh ( as/2) \cosh (a p/2 ) - 2 i \varepsilon \cosh (a s/2 ) \cosh (ap/2) \Big)^2 
    \non 
    \\
    & \qquad - \Big( 2a^{-1} \sinh (as/2 ) \sinh (ap/2) +  l - 2 i \varepsilon \cosh (as /2 ) \sinh (ap/2 ) \Big)^2 \bigg]^{-1} .
\end{align}
The condition for a pole is given by:
\begin{align}
    0 &= \Big( 2a^{-1} \sinh(as/2) \cosh(ap/2) - 2i \varepsilon \cosh(as/2) \cosh(ap/2) \Big)^2 
    \non 
    \\
    & \qquad - \Big( 2a^{-1} \sinh(as/2) \sinh(ap/2) - 2i \varepsilon \cosh(as/2) \sinh(ap/2) + l \Big)^2 .
\end{align}
We absorb terms into the infinitesimal regularisation constant $\varepsilon$:
\begin{align}
    0 &= \Big( 2a^{-1} \sinh(as/2) \cosh(ap/2) - i \varepsilon \Big)^2 - \Big( 2a^{-1} \sinh(as/2) \sinh(ap/2) - i \varepsilon + l \Big)^2 ,
    \non 
    \\
    0 &= 4 a^{-2} \sinh^2(as/2) \cosh^2(ap/2) - 4i \varepsilon a^{-1} \sinh(as/2) \cosh(ap/2) - \varepsilon^2 
    \vphantom{\Big)}
    \non 
    \\
    & \qquad - 4a^{-2} \sinh^2(as/2) \sinh^2(ap/2) - 4la^{-1} \sinh(as/2) \sinh(ap/2) 
    \non 
    \vphantom{\Big)}
    \\
    & \qquad + 4i \varepsilon a^{-1} \sinh(as/2) \sinh(ap/2) - l^2 + 2i \varepsilon l + \varepsilon^2 \vphantom{\Big)} ,
    \non 
    \\
    0 &= 4a^{-2} \sinh^2(as/2) - 4a^{-1} \sinh(as/2) \Big( i \varepsilon  e^{-ap/2} + l \sinh(ap/2 ) \Big) - l^2 + 2 i \varepsilon l .
\intertext{Further simplifying, we obtain:}
    0 &= 4a^{-2} \sinh^2(as/2) - 4a^{-1} \sinh(as/2) \Big( i \varepsilon + l \sinh(ap/2) \Big) - l^2 + i \varepsilon ,
    \non 
    \\
    0 &= 4a^{-2} \sinh^2(as/2) - 4i \varepsilon a^{-1} \sinh(as/2) - 4a^{-1} l \sinh(as/2) \sinh(ap/2) - l^2 + i \varepsilon ,
    \vphantom{\Big)} 
    \non 
    \\
    0 &= 4a^{-2} \sinh^2(as/2) - 4a^{-1} l \sinh(as/2) \sinh(ap/2) - l^2 + i \varepsilon 
    \vphantom{\Big)} .
\end{align}
In evaluating the transition probability, we shift the contour of the $s$ integration variable in the complex plane. We thus isolate $s$ and $p$ on either side of the equation:
\begin{align}
    4a^{-1} l  \sinh(as/2) \sinh(ap/2) &= 4a^{-2} \sinh^2(as/2) - l^2 + i \varepsilon 
    \vphantom{\Big) }
    \non 
    \\
    l \sinh(as/2) \sinh(ap/2) &= a^{-1} \sinh^2(as/2) - a l^2/4 + i \varepsilon 
    \vphantom{\Big) }
    \non 
    \\
    \sinh(ap/2) &= ( a l )^{-1} \sinh(as/2) - (a l/4) \mathrm{csch}(as/2) + i \varepsilon 
    \vphantom{\Big) } .
\end{align}
Let $s = x + i y$ be a complex variable (of which we perform the contour shift in the complex plane). Then, we take the imaginary part of the resulting equation to obtain a constraint on the allowed regimes for this shift:
\begin{align}
    0 &= (al)^{-1} \cosh(ax/2) \sin(ay/2) - (al/4) \frac{2\cosh(ax/2) \sin(ay/2)}{\cos(ay) - \cosh(ax)} + \varepsilon ,
    \non 
    \\ 
    \label{eq8.82}
    0 &= \cosh(ax/2) \sin(ay/2) \left( (al)^{-1} - \frac{al}{2} \frac{1}{\cos(ay) - \cosh(ax)} \right) + \varepsilon .
\end{align}
There are two cases in which Eq.\ ( \ref {eq8.82}) might be satisfied. The first is when $\sin(ay/2) =  0$, which motivates us to consider only those regimes in which 
\begin{align}
    a \sigma^2 \Omega &< \pi 
    \vphantom{\frac{\pi}{a}} 
\end{align}
(since we are shifting the contour by $ 2\sigma^2 \Omega$ in the complex plane). Meanwhile the term in the brackets (in the case when $x = 0$, since we assume the saddle-point approximation in which $\mathrm{Re} (s ) = 0$ i.e.\ for a narrowly peaked Gaussian switching function) is zero when:
\begin{align}
    \frac{1}{al} &= \frac{al}{2} \frac{1}{\cos(ay) - 1} ,
\end{align}
which is never satisfied, and as desired. A similar procedure may be applied to the other Wightman function, yielding an identical constraint for the valid regime of analysis.

\subsection{Pole Constraints: Antiparallel Accelerations}
In the antiparallel case, the Wightman function is 
\begin{align}
    W_{12}(s,p) &= - \frac{1}{4\pi^2} \bigg[ \Big( 2\cosh(ap/2) \Big( a^{-1} \sinh( as/2) - i\varepsilon \cosh (as/2) \Big) \Big)^2 
    \non 
    \\
    &\qquad - \Big( 2 \cosh (ap/2) \Big( a^{-1} \cosh(as/2) - i\varepsilon \sinh(as/2) \Big) - 2a^{-1} + l \Big)^2 \bigg]^{-1}. 
\end{align}
The pole condition is thus,
\begin{align}
    0 &= \Big( 2 \cosh(ap/2) \Big(a^{-1} \sinh(as/2) - i\varepsilon \cosh(as/2) \Big) \Big)^2 
    \non 
    \\
    & \qquad - \Big( 2 \cosh(ap/2) \Big( a^{-1} \cosh(as/2) - i\varepsilon \sinh(as/2) \Big) - 2 a^{-1} + l \Big)^2 ,
    \non 
    \\
    &= \Big( 2a^{-1} \cosh(ap/2) \sinh(as/2) - 2i \varepsilon \cosh(ap/2) \cosh(as/2) \Big)^2 
    \non 
    \\
    & \qquad - \Big( 2 a^{-1} \cosh(ap/2) \cosh(as/2) - 2 i \varepsilon \cosh(ap/2) \sinh(as/2) - 2a^{-1} + l \Big)^2 .
\intertext{As before, we can absorb various terms, without loss of generality, into the infinitesimal constant $\varepsilon$:}
    0 &= \Big( 2a^{-1} \cosh(ap/2) \sinh(as/2) - i \varepsilon \Big)^2 - \Big( 2a^{-1} \cosh(ap/2) \cosh(as/2) - i \varepsilon - 2a^{-1} + l \Big)^2,
    \non 
    \\
    0 &= - 4 a^{-2} \cosh^2(ap/2) + \cosh(ap/2) \cosh(as/2) \Big( 8a^{-2} - 4a^{-1}l \Big) - \left( l - 2 a^{-1} \right)^2 + i \varepsilon .
\end{align}
Isolating $p$ and $s$:
\begin{align}
    \cosh( ap/2) \cosh(as/2) \left( 8a^{-2} - 4a^{-1}l \right) &= 4a^{-2} \cosh^2(ap/2) + \left( l - 2a^{-1} \right)^2 - i \varepsilon 
    \vphantom{\frac{\left( l - 2a^{-1} \right)^2}{8a^{-2} - 4a^{-1} l}}
    \non 
    \\
    \cosh(as/2) \left( 8a^{-2} - 4a^{-1} l \right) &= 4a^{-2} \cosh(ap/2) + \left( l - 2a^{-1} \right)^2 \mathrm{sech}(ap/2) - i \varepsilon 
    \vphantom{\frac{\left( l - 2a^{-1} \right)^2}{8a^{-2} - 4a^{-1} l}}
    \non 
    \\
    \cosh(as/2) &= \frac{4a^{-2}}{8a^{-2} - 4a^{-1}l} \cosh(ap/2) + \frac{\left( l - 2a^{-1} \right)^2}{8a^{-2} - 4a^{-1} l} \mathrm{sech}(ap/2) - i \varepsilon \non 
\end{align}
As before, we let $s = x + iy$ and taking the imaginary part of the equation, yielding:
\begin{align}
    0 &= \sinh(ax/2) \sin(ay/2) + i \varepsilon .
\end{align}
This implies the same constraint as the parallel superposition case.

\section{Calculation of the Visibility}
In this section, we derive the interferometric visibility of the control superposition. We focus on the case of a two-trajectory superposition, although the calculation easily generalises to $N$ amplitudes. We have the initial state of the control, field, and detector given by 
\begin{align}
    | \psi \rangle &= \frac{1}{\sqrt{2}} \left( | 1 \rangle + | 2 \rangle \right) | 0_M \rangle | g \rangle ,
\end{align}
where the states $\{ | 1 \rangle, | 2 \rangle\}$ denote the trajectory states of the detector. Evolving this initial state in time yields, 
\begin{align}
    \hat{U} | \psi \rangle &= \frac{1}{\sqrt{2}} \left( \hat{U}_1 | 1 \rangle + \hat{U}_2 | 2 \rangle \right) | 0_M \rangle | g \rangle .
\end{align}
It is convenient to work with the density matrix:
\begin{align}
    \hat{\rho}_\mathrm{CFD} \equiv \hat{U} | \psi \rangle \langle \psi | \hat{U}^\dd &= \frac{1}{2} \Big( \hat{U}_1 |1 \rangle\langle 1 | \otimes \hat{\rho}_\mathrm{FD}  \hat{U}_1^\dd + \hat{U}_1 | 1 \rangle\langle 2 | \otimes \hat{\rho}_\mathrm{FD} \hat{U}_2^\dd 
    \non 
    \\
    & \qquad + \hat{U}_2 | 2 \rangle\langle 1  | \otimes \hat{\rho}_\mathrm{FD} \hat{U}_1^\dd + \hat{U}_2 | 2 \rangle\langle 2 | \otimes \hat{\rho}_\mathrm{FD} \hat{U}_2^\dd \Big) ,
\end{align}
where $
\hat{\rho}_\mathrm{FD} = | 0_M\rangle\langle 0_M | \otimes | g \rangle\langle g |$. Tracing out the field and detector degrees of freedom gives the individual terms, 
\begin{align}
    \mathrm{Tr}_{\Phi,D} \big[ \hat{\rho}_{ij,\mathrm{CFD}} \big] &=
    \frac{1}{2} \left( 1 + P_{ij,E} - \frac{P_{ii,E} + P_{jj,E}}{2} \right),
\end{align}
where
\begin{align}
    P_{ij,E} &= \lambda^2 \infint\D \tau \infint\D \tau' \eta(\tau) \eta(\tau') e^{-i\Omega(\tau - \tau')} W(\Mtx_i(\tau) , \Mtx_j'(\tau') ) 
\end{align}
as usual. The state of the control after the interaction is thus:
\begin{align}
    \mathrm{Tr}_{\Phi,D} \big[ \hat{\rho}_\mathrm{CFD} \big] &= \frac{|1\rangle\langle 1 |}{2} \left( 1 + P_{11,E} - P_{11,E} \right) + \frac{|1 \rangle\langle 2 | }{2} \left( 1 + P_{21,E} - \frac{P_{11,E} + P_{22,E}}{2} \right) 
    \non 
    \\
    & \qquad + \frac{|2\rangle\langle 1|}{2} \left( 1 + P_{12,E} - \frac{P_{11,E} + P_{22,E}}{2} \right) + \frac{|2 \rangle\langle 2|}{2} \left( 1 + P_{22,E} - P_{22,E} \right) .
\end{align}
We now measure the control in the superposition basis $| \nu \rangle = ( | 1\rangle + \exp(i \phi ) | 2 \rangle ) /\sqrt{2}$ with variable phase $\phi$:
\begin{align}
    \left\langle \nu \left| \mathrm{Tr}_{\Phi,D} \big[ \hat{\rho}_\mathrm{CFD} \big] \right| \nu\right\rangle &= \frac{1}{4} \left( 1 + P_{11,E} - P_{11,E} \right) + \frac{e^{i\phi}}{4} \left( 1+ P_{21,E} - \frac{P_{11,E} + P_{22,E}}{2} \right) 
    \non 
    \\
    & \qquad + \frac{e^{-i\phi}}{4} \left( 1 + P_{12,E} - \frac{P_{11,E} + P_{22,E}}{2} \right) + \frac{1}{4} \left( 1 + P_{22,E} - P_{22,E} \right) , \non 
    \\
    &= \frac{1}{2} \left( 1 + \left( 1 + P_{12,E} - \frac{P_{11,E} + P_{22,E}}{2} \right) \cos(\delta \phi) \right) .
\end{align}
The coefficient of the $\cos(\delta\phi)$ term, as stated previously, is known as the interferometric visibility, and quantifies the amount of coherence remaining in the control. This is the result stated in the main text.

\end{widetext}

\bibliography{References.bib}

\end{document}